\documentclass[aps,prd,preprint,groupedaddress,amsmath,amssymb]{revtex4}

\begin{document}


\title{Weisskopf-Wigner Decay Theory for the Energy-Driven Stochastic Schr\"odinger Equation}


\author{Stephen L. Adler}
\email[]{adler@ias.edu}
\affiliation{Institute for Advanced Study\\
Einstein Drive\\
Princeton, NJ 08540}


\date{September, 2002}
\begin{abstract}
We generalize the Weisskopf-Wigner theory for the line shape and transition 
rates of decaying states to the case of the energy-driven stochastic 
Schr\"odinger equation that has been used as a 
phenomenology for state vector reduction.  Within the standard approximations  
used in the Weisskopf-Wigner analysis, and assuming that the perturbing 
potential inducing the decay has vanishing matrix elements within the 
degenerate manifold containing the decaying state, the stochastic 
Schr\"odinger equation linearizes. Solving the linearized equations,  
we find no change from the standard analysis in the line shape 
or the transition rate per unit time.  The only effect of the stochastic 
terms is to alter the early time transient behavior of the decay, in a way 
that eliminates the quantum Zeno effect.  We apply our results to estimate 
experimental bounds on the parameter governing the stochastic effects.  In an Added Note, elegant stochastic-theoretic methods suggested by Di\'{o}si are used to rederive the principal results, without the assumptions needed to linearize the stochastic equation, and to give analogous results for the Rabi oscillations of a two-level system. 

\end{abstract}


\maketitle


\section{\label{jam}Introduction}

There has recently been considerable interest in the possibility that 
quantum mechanics, and the Schr\"odinger equation, may be modified 
at a very low level by effects arising from Planck scale physics.  Such 
speculations have been motivated on the one hand by considerations from 
string theory \cite{ellis} and quantum gravity \cite{penrose}, and on the other hand by  
efforts \cite{penrose,pearle,gisin} to achieve an objective equation describing state vector reduction. The majority of the objective reduction discussions fall 
into two classes:  those 
that postulate a stochastic process producing spatial localization \cite{pearle}, and 
those that postulate an analogous stochastic process leading to localization 
in energy \cite{gisin} (the so-called ``energy-driven'' stochastic Schr\"odinger   
equation.)  
Both the spatial localization  
and the energy localization stochastic Schr\"odinger equations  
avoid problems with superluminal 
signal propagation that characterize attempts at 
deterministic nonlinear 
modifications of the Schr\"odinger equation \cite{bialy}.  
We find the energy-driven approach particularly  
appealing because it is energy conserving, leads with no approximations 
to Born rule probabilities and to the L\"uders projection postulate, 
has sensible clustering properties, and when environmental interactions are 
taken into account explains state vector reduction with a single Planck scale 
stochastic parameter \cite{gisin,adler}.   

Although physical prejudices might suggest a Planck scale magnitude for the 
stochastic parameter in the energy-driven equation, 
one can instead take the point of view that the 
stochastic parameter can have {\it a priori} any value, and use 
current experimental information to place bounds on it.  This  
approach has been pursued \cite{lisi} in the context of particle physics systems that  
exhibit oscillations between different mass eigenstates (the $K$-meson, 
$B$-meson, and neutrino systems), with results that are summarized in the 
final  section of this paper.  An alternative source of bounds on the 
stochastic parameter could come from experiments observing decays and line 
shapes in atomic and particle systems, since if the stochastic terms in 
the Schr\"odinger equation were to change
the standard Weisskopf-Wigner analysis of decay processes 
in a significant way, then observable effects could result.   
Thus, to pursue phenomenological studies of the energy-driven 
equation, it is important to generalize the standard Weisskopf-Wigner decay  
theory \cite{weiss} to include effects of the energy-driven stochastic terms.  This is 
the problem that is analyzed in this paper.  

\section{\label{jell}The Energy-Driven Stochastic Schr\"odinger Equation and Properties of the 
It\^o Stochastic Calculus}

Letting $|\psi\rangle$ denote a unit normalized Schr\"odinger 
picture state vector, 
the standard form \cite{pearle,gisin,adler} of the energy-driven stochastic 
Schr\"odinger equation is (with $\hbar=1$) 
\begin{subequations}
\label{onemulti}
\begin{equation}
d|\psi\rangle=-iH|\psi\rangle dt - \frac{1}{8} \sigma^2
(H-\langle H \rangle)^2 |\psi\rangle dt + \frac{1}{2} \sigma 
(H-\langle H \rangle) |\psi\rangle  dW_t~~~.
\label{amber}   
\end{equation}
Here $H$ is the Hamiltonian, $\langle H \rangle=\langle \psi|H|\psi \rangle$ 
is the expectation of the Hamiltonian in the state $|\psi\rangle$, 
$\sigma$ is a 
numerical parameter governing the strength of the stochasticity, 
and $dW_t$ is 
an It\^o stochastic differential that, together with $dt$, obeys the standard 
It\^o calculus rules \cite{gardiner} 
\begin{equation}
dW_t^2=dt~,~~dW_t dt =dt^2=0~~~.
\label{beige}  
\end{equation}
\end{subequations}
By construction, the nonlinear evolution of 
Eq.~(\ref{amber}) guarantees the preservation in time of the unit 
normalization of the state vector $|\psi \rangle$.  

In the following sections, we shall need a number of properties of the 
It\^o calculus that we summarize here.  First of all, in the It\^o calculus 
the Leibnitz chain rule generalizes to 
\begin{subequations}
\label{twomulti}
\begin{equation}
d(AB) = (A+dA)(B+dB)-AB= (dA) B + A dB + dA dB~~~,
\label{almond}  
\end{equation}
with the final term $dA dB$  contributing a term proportional 
to $dt$ when the $dW_t$ contributions to both $dA$ and $dB$ are nonzero.  
Letting $W_t$ be the Brownian motion 
\begin{equation}
W_t = \int_0^t dW_u ~~~,
\label{bean} 
\end{equation}
we see in particular that
\begin{equation} 
d\exp(\alpha W_t)= \exp(\alpha W_t)[\exp(\alpha dW_t)-1]
=\exp(\alpha W_t)[\alpha dW_t + \frac{1}{2} \alpha^2 dt]~~~.
\label{cocoa} 
\end{equation}
\end{subequations}
Letting $E[...]$ denote the stochastic expectation of its argument, 
and letting $A(t)$ denote any function of the stochastic 
process up to time $t$, 
we have
\begin{subequations}
\label{threemulti}
\begin{equation} 
E[dW_t A(t)]=0~~~,
\label{alpha} 
\end{equation}
since the It\^o differential refers to the 
time interval from $t$ to $t+dt$, and hence is statistically independent of 
the process up to time $t$.  Thus, taking the expectation of Eq.~(\ref{cocoa}), 
we get the differential equation
\begin{equation} 
dE[\exp(\alpha W_t)]= E[\exp(\alpha W_t)] \frac{1}{2} \alpha^2 dt~~~,
\label{beta} 
\end{equation}
which can be immediately integrated to give
\begin{equation} 
E[\exp(\alpha W_t)]=\exp(\frac{1}{2} \alpha^2 t)~~~,
\label{chi}  
\end{equation}
\end{subequations}
a result that will be needed later on.

Let us make an elementary application of the It\^o formalism, to 
write the stochastic Schr\"odinger equation of Eq.~(\ref{amber}) in 
an equivalent form.  First of all, forming the density matrix 
\begin{subequations}
\label{fourmulti}
\begin{equation}
\rho=|\psi \rangle \langle \psi|~~~,
\label{azure}  
\end{equation}
we have from Eq.~(\ref{almond}),
\begin{equation} 
d\rho =(d|\psi \rangle) \langle \psi| + |\psi \rangle d\langle \psi|
+d|\psi \rangle d \langle \psi|  ~~~, 
\label{blue} 
\end{equation} 
which on substitution of Eq.~(\ref{amber}) and use of the  It\^o calculus rules of 
Eq.~(\ref{beige}) gives the evolution equation for the density matrix, 
\begin{equation}
d\rho =  i[\rho, H]dt -\frac{1}{8}\sigma^2 [H,[H,\rho]] dt
+ \frac{1}{2} \sigma [\rho,[\rho,H]] dW_t~~~.
\label{clear} 
\end{equation}
\end{subequations}
Taking the stochastic expectation of this equation, using Eq.~(\ref{alpha}), gives 
a differential equation of the Lindblad type \cite{lind} for $E[\rho]$, 
\begin{equation}
\frac{dE[\rho]}{dt}= i[E[\rho],H] - \frac{1}{8} \sigma^2 [H,[H,E[\rho]]]~~~.
\label{deep}  
\end{equation}
The fact that this equation is linear (in contrast to Eq.~(\ref{clear}), which is 
nonlinear) is the fundamental  
reason \cite{bialy} why Eq.~(\ref{amber}) does not give rise to superluminal 
signal propagation.  
 
\section{\label{jet}Initial Formulation of the Decay Problem}

Let us now formulate the decay problem for the stochastic Schr\"odinger 
equation of Eq.~(\ref{amber}), following the standard procedure for the    
usual Schr\"odinger equation without stochastic terms.  
We suppose that for times $t \leq 0$ the 
Hamiltonian $H$ is given by an unperturbed Hamiltonian $H_0$, with  
eigenstates $|n\rangle$ and eigenvalues $E_n$, 
\begin{subequations}
\label{sixmulti}
\begin{equation}
H_0 |n\rangle = E_n |n\rangle~~~, 
\label{april} 
\end{equation}
and that the system under consideration is in an eigenstate $|s_A\rangle$
with eigenvalue $E_s$, which is one of a set of degenerate energy eigenstates 
$|s_a\rangle~,~~a=1,...,D$.  Because Eq.~(\ref{amber}) acts as an ordinary 
Schr\"odinger evolution on a state $|\psi\rangle$ that is an energy 
eigenstate, the system remains in the state $|s_A\rangle$ as long as the 
Hamiltonian remains equal to $H_0$.  Hence the starting point for the 
standard decay analysis \cite{merz} is also a consistent starting point for its 
stochastic extension under Eq.~(\ref{amber}).  As in the standard procedure, we 
assume that at $t=0$ a time-independent perturbation $V$ is switched on, so 
that for times $t>0$ the Hamiltonian is $H=H_0+V$.  The initial state 
$|s_A\rangle$ is then no longer an energy eigenstate, and so will decay 
into various other states $|m\rangle$; our problem, as in the usual case, 
is to find the partial transition rates 
for this decay and the probability amplitude 
for the system to remain in the initial degenerate group of states.    

In formulating this problem, it is convenient to expand the state 
$|\psi\rangle$ over the basis $|n\rangle$ and, as in the standard case, 
to remove the Schr\"odinger time evolution associated with the unperturbed 
Hamiltonian $H_0$, by writing
\begin{equation}
|\psi(t)\rangle = \sum_n |n\rangle \exp(-iE_nt) C_n(t)~~~.
\label{betsy}  
\end{equation}
\end{subequations}
Substituting Eq.~(\ref{betsy}) into Eq.~(\ref{amber}), and projecting on $\langle m|$, it 
is a matter of straightforward but somewhat tedious algebra to compute 
the stochastic evolution equation for $C_m(t)$, with the result 
\begin{subequations}
\label{sevenmulti}
\begin{eqnarray}
dC_m(t)&=&\alpha_mC_m(t)+ \sum_n \beta_{mn} C_n(t)~~~,\nonumber\\
\alpha_m&=&-\frac{1}{8} \sigma^2 (E_m-\langle H \rangle)^2 dt 
+\frac{1}{2} \sigma (E_m-\langle H \rangle) dW_t~~~,\nonumber\\
\beta_{mn}&=&-iV_{mn}\exp[i(E_m-E_n)t]dt
-\frac{1}{8} \sigma^2[(E_m+E_n-2\langle H \rangle)V_{mn} + (V^2)_{mn}]
\exp[i(E_m-E_n)t]dt\nonumber\\
&+&\frac{1}{2} \sigma V_{mn}\exp[i(E_m-E_n)t]dW_t~~~.
\label{ablaze}  
\end{eqnarray}
The corresponding expression for $\langle H \rangle$ is 
\begin{equation}
\langle H \rangle = \sum_n E_n |C_n(t)|^2 + \sum_{mn} V_{mn} 
\exp[i(E_m-E_n)t] C_m^*(t) C_n(t)~~~.
\label{baggy}  
\end{equation}
In these equations $V_{mn}$ and $(V^2)_{mn}$ denote the respective 
matrix elements 
\begin{equation}
V_{mn}=\langle m|V| n \rangle~,~~(V^2)_{mn}= \langle m|V^2| n \rangle
~~~. 
\label{cache} 
\end{equation}
\end{subequations}

\section{\label{joke}Approximation to Leading Order in $V$}

Equations (\ref{sevenmulti}a-c) are a complicated, nonlinear set of stochastic differential 
equations, and so to solve them approximations will be needed.  Following the 
Weisskopf-Wigner analysis, we shall make the approximation of regarding 
$V$ as a small perturbation.  The coefficients $C_m,~m\not\in \{s_a\}$ for  
states not in the initial degenerate manifold will then be of order 
$O(V)$, and we neglect $O(V^2)$ and higher contributions to them   
(except those arising implicitly through our solution for the $C_{s_a}$).  
On the other hand, the coefficients $C_{s_a}$ of states in the degenerate 
manifold can be of order unity, and we calculate these coefficients to 
order $V^2$ accuracy, neglecting corrections of order $V^3$ and higher.  
In a similar fashion, in expressions involving the stochasticity parameter 
$\sigma$, we shall retain terms of order $\sigma V$ and its powers 
$(\sigma V)^2$, etc., but shall neglect terms of order $\sigma V^2$ and 
higher that involve extra factors of $V$ relative to the terms that we 
are retaining.  Finally, although we shall see that 
$E_m-E_s$ is effectively small, 
we shall retain all terms of order 
$\sigma (E_m-E_s)$, $\sigma^2(E_m-E_s)$, $[\sigma (E_m-E_s)]^2$, etc., but 
shall drop terms $\sigma (E_m-E_s)\sigma O(V^2)$ that are 
smaller than these by a factor of order $\sigma V^2$ or $V^2$.

Making use of the perturbative ordering of the coefficients $C_n$, 
we begin by simplifying and approximating the expression in Eq.~(\ref{baggy}) 
for $\langle H \rangle$.  Separating off the states in the initial degenerate 
manifold, the sum in Eq.~(\ref{baggy}) becomes 
\begin{subequations}
\label{eightmulti}
\begin{equation}
\langle H \rangle = E_s \sum_a|C_{s_a}|^2  
+ \sum_{ab} V_{s_as_b}C^*_{s_a}C_{s_b} + O(V^2)~~~.
\label{acadia} 
\end{equation}
However, since the state vector $|\psi\rangle$ remains unit normalized, 
we have 
\begin{equation}
\sum_a|C_{s_a}|^2 =1 - \sum_{m\not\in\{s_a\}}|C_m|^2 =1+O(V^2)~~~,
\label{beech} 
\end{equation}
and so we have 
\begin{equation}
\langle H \rangle = E_s   
+ \sum_{ab} V_{s_as_b}C^*_{s_a}C_{s_b} + O(V^2)~~~.
\label{cedar}  
\end{equation}
\end{subequations}
     
If we substitute Eq.~(\ref{cedar}) back into Eq.~(\ref{ablaze}), we are still left with 
a nonlinear set of equations.  Therefore we shall 
also introduce the simplifying assumption  
that the perturbing potential has vanishing matrix elements within the 
degenerate manifold containing the initial state,  so that 
\begin{subequations}
\label{ninemulti}
\begin{equation}
V_{s_as_b}=0~,~~a,b=1,...,D~~~.
\label{acrid} 
\end{equation}
There are important, physically relevant cases in which Eq.~(\ref{acrid}) is obeyed 
as a result of selection rules.  For example, for radiative 
decays treated in the electric dipole approximation, with $H_0$ taken as 
the atomic Hamiltonian plus the free radiation Hamiltonian, 
and with $V$ taken as the atomic coupling to the transverse 
electromagnetic modes, Eq.~(\ref{acrid}) is obeyed 
as a result 
of parity invariance when the states in the initial degenerate manifold 
all have the same parity.  (We caution, however, that 
Eq.~(\ref{acrid}) is not valid for the analysis 
of $K$ (or $B$) meson systems when $H_0$ is taken as the strong interaction 
Hamiltonian. Thus here either one has to employ the nonlinear 
equations following from Eq.~(\ref{cedar}), or one has to redefine 
$H_0$ so as to impose Eq.~(\ref{acrid}) by including in $H_0$ the  
$|\Delta S|=2$ (or $|\Delta C|=2$) weak interaction effective Hamiltonian 
terms, with $V$ defined 
to contain only the $|\Delta S|=1$ (or $|\Delta C|=1$) weak interaction 
terms responsible for $K$ (or $B$) meson decays. Such a redefinition is 
consistent in the vacuum saturation approximation for the 
$|\Delta S|=2$ (or $|\Delta C|=2$) terms.)   

With the simplifying assumption of Eq.~(\ref{acrid}), Eq.~(\ref{cedar}) becomes simply 
\begin{equation}
\langle H \rangle = E_s   +  O(V^2)~~~.
\label{bevel}  
\end{equation}
\end{subequations}
Substituting this into Eq.~(\ref{ablaze}), and dropping terms that are not of leading 
order in $V$ in the sense defined above, Eqs.~(\ref{ablaze},\ref{baggy}) simplify to the   
following set of linear equations, 
\begin{subequations}
\label{tenmulti}
\begin{eqnarray}
dC_m(t)&=&(\alpha_m^{(1)} dW_t + \alpha_m^{(2)} dt) C_m(t)   
+\sum_n 
\exp[i(E_m-E_n)t](\gamma_{mn}^{(1)} dW_t 
+ \gamma_{mn}^{(2)} dt) C_n(t) ~~~,\nonumber \\
\alpha_m^{(1)}&=&{1\over 2} \sigma (E_m-E_s)~,~~
\alpha_m ^{(2)}=-{1\over 8}\sigma^2(E_m-E_s)^2=-{1\over 2} (\alpha_m^{(1)})^2 ~~~,\nonumber \\
\gamma_{mn}^{(1)}&=&{1\over 2} \sigma V_{mn}~,~~
\gamma_{mn}^{(2)}=-iV_{mn}-{1\over 8} \sigma^2[(E_m+E_n-2E_s)V_{mn}
+(V^2)_{mn}]~~~.
\label{adieu}  
\end{eqnarray}
Corresponding to the magnitude ordering of the coefficients $C_m$ introduced 
above, it is convenient to rewrite Eq.~(\ref{adieu}) as separate equations 
for the two cases, $m\in \{s_a\}$ and 
$m \not\in \{s_a\}$.  For $m\in \{s_a\}$ the coefficients $\alpha_s^{(1,2)}$
vanish; separating the sum over $n$ into terms where $n\in \{s_a\}$  and 
$n \not\in \{s_a\}$, using the assumption of Eq.~(\ref{acrid}), and dropping terms 
of nonleading order in $V$, we get
\begin{eqnarray}
dC_{s_a}(t)&=&-{1\over 8} \sigma^2 dt \sum_b (V^2)_{{s_a}{s_b}}C_{s_b}(t) \nonumber \\
&+&\sum_{n \not\in \{s_a\} }
\exp[i(E_s-E_n)t](\gamma_{s_an}^{(1)} dW_t 
+ \gamma_{s_an}^{(2)} dt) C_n(t) ~~~,\nonumber \\
\gamma_{s_an}^{(1)}&=&{1\over 2} \sigma V_{s_an}~,~~
\gamma_{s_an}^{(2)}\simeq -iV_{s_an}f_n~~~,
\label{bijou} 
\end{eqnarray} 
where we have introduced the definition
\begin{equation} 
f_n=1-{i\over 8} \sigma^2(E_n-E_s)~~~.
\label{canape} 
\end{equation}
For $m \not\in \{s_a\}$  the coefficients $\alpha_m^{(1,2)}$ are nonzero, 
but only the terms with $n\in \{s_a\}$ have to be retained in the sum over 
$n$, and so we similarly get 
\begin{eqnarray}
dC_m(t)&=&(\alpha_m^{(1)} dW_t + \alpha_m^{(2)} dt) C_m(t)  \nonumber \\ 
&+&\exp[i(E_m-E_s)t] \sum_a(\gamma^{(1)}_{ms_a} dW_t+\gamma^{(2)}_{ms_a}dt)
C_{s_a}(t)~~~,\nonumber \\
\gamma_{ms_a}^{(1)}&=&{1\over 2}\sigma V_{ms_a}~,~~
\gamma_{ms_a}^{(2)}\simeq -iV_{ms_a}f_m~~~.
\label{devotee}   
\end{eqnarray}
\end{subequations}
Equations (\ref{tenmulti}a-d) are the basic system of stochastic differential 
equations that we shall solve in the subsequent sections.  
\bigskip
\section{\label{juror}Equations for Expectations of the Coefficients}

The principal quantities that we wish to calculate are the expectations  
$E[|C_m(t)|^2]$ of the squared magnitudes of the coefficients, since these 
give the expectations of the probabilities for the various states to be 
occupied.  We shall show in this section that, within our approximations, 
these can be directly related to the expectations $E[C_m(t)]$ of the 
coefficients themselves, for which we shall derive a closed, linear set 
of ordinary differential equations.  

Again, we consider separately the cases $m\in\{s_a\}$ 
and $m\not\in\{s_a\}$.  For $C_{s_a}$, we write 
\begin{subequations}
\label{elevenmulti}
\begin{equation}
C_{s_a}(t)=E[C_{s_a}(t)]+ \Delta_a(t)~~~, 
\label{africa}  
\end{equation}
with $E[\Delta_a(t)]=0$, and with $\Delta_a(0)=0$ since the stochastic 
terms in the differential equation act only after $t=0$.  However, referring 
to Eq.~(\ref{bijou}) we see that $dC_{s_a}$ is of order $V^2$, and so 
$\Delta_a(t)$ must also be of order $V^2$. Therefore
\begin{equation}  
E[|C_{s_a}(t)|^2]=|E[C_{s_a}(t)]|^2 +O(V^4)~~~,
\label{brazil}  
\end{equation}
\end{subequations}
and so to the accuracy to which we are working, we can compute 
$E[|C_{s_a}(t)|^2]$ from the expectation $E[C_{s_a}(t)]$, ignoring the 
effects of fluctuations.  

We consider next $E[|C_m(t)|^2]$ for $m\not\in\{s_a\}$. Applying the It\^o 
rule of Eq.~(\ref{almond}), we have
\begin{subequations}
\label{twelvemulti}
\begin{equation} 
dE[|C_m(t)|^2]=E[(dC_m^*(t)) C_m(t) + C_m^*(t) dC_m(t) + dC_m^*(t) dC_m(t)] ~~~.
\label{agate}  
\end{equation}
Substituting Eq.~(\ref{devotee}) for $dC_m(t)$ and using Eq.~(\ref{alpha}), which eliminates 
the $dW_t$ contributions, and using the fact that to leading order in $V$ 
we can replace $C_{s_a}(t)$ by its expectation, we get after some algebraic 
simplification the formula  
\begin{eqnarray}
{d \over dt}E[|C_m(t)|^2]  
&=&\exp[-i(E_m-E_s)t]if_mE[C_m(t)]\sum_aV_{ms_a}^*E[C_{s_a}^*(t)] \nonumber \\  
&-&\exp[ i(E_m-E_s)t]if_m^*E[C_m^*(t)]\sum_aV_{ms_a}E[C_{s_a}(t)] \nonumber \\  
&+&{1\over 4} \sigma^2 |\sum_a V_{ms_a}E[C_{s_a}(t)]|^2~~~,
\label{brooch} 
\end{eqnarray}
\end{subequations}
which can be integrated to give $E[|C_m(t)|^2]$ once the expectations 
$E[C_m(t)]$ and $E[C_{s_a}(t)]$ are known.  

To get a closed set of equations for the expectations of the coefficients, 
we simply take the expectations of Eqs.~(\ref{bijou}) and (\ref{devotee}), and use Eq.~(\ref{alpha}), 
which again eliminates the $dW_t$ contributions.  For $E[C_{s_a}]$ we thus 
get
\begin{subequations}
\label{thirteenmulti} 
\begin{eqnarray}
{d\over dt}E[C_{s_a}(t)]&=&-{1\over 8} \sigma^2  \sum_b (V^2)_{{s_a}{s_b}}
E[C_{s_b}(t)] \nonumber \\
&+&\sum_{n \not\in \{s_a\} }
\exp[i(E_s-E_n)t] (-i) V_{s_an}f_n E[C_n(t)] ~~~,
\label{algae} 
\end{eqnarray}
while for $E[C_m(t)]$ with $m\not \in \{s_a\}$ we find
\begin{eqnarray}
{d\over dt}E[C_m(t)]&=& -{1\over 8}\sigma^2(E_m-E_s)^2) E[C_m(t)]  \nonumber \\ 
&+&\exp[i(E_m-E_s)t] \sum_b(-i)V_{ms_b}f_m E[C_{s_b}(t)]~~~.
\label{botany} 
\end{eqnarray}
\end{subequations}

\section{\label{jest}Solutions for Expectations of the Coefficients}

We proceed now to solve the linear system of equations for the expectations 
of the coefficients given in Eqs.~(\ref{thirteenmulti}a,b).  Since the problem is defined 
on the half line $t>0$, the natural way to do this is by using the 
Laplace transform.   Defining 
\begin{subequations}
\label{fourteenmulti}
\begin{equation}
g_m(p)=\int_0^{\infty}dt \exp(-pt)  E[C_m(t)]~~~,
\label{auntie}  
\end{equation}
we have, by an integration by parts, 
\begin{equation}
\int_0^{\infty}dt \exp(-pt) {d E[C_m(t)]\over dt} 
=pg_m(p)-E[C_m(0)]~~~,
\label{blanket}  
\end{equation}
and also
\begin{equation}
\int_0^{\infty}dt \exp(-pt) \exp[i(E_m-E_n)t] E[C_n(t)]=g_n(p-iE_m+iE_n)~~~,
\label{cover}  
\end{equation}
with the integrals in Eqs.~(\ref{fourteenmulti}a-c) defining analytic functions of $p$ in the 
right hand half plane ${\rm Re}\,p>0$.  The inversion of the Laplace transform 
is given by the formula
\begin{equation}
E[C_m(t)]={1 \over 2\pi i}\int_{\epsilon-i\infty}^{\epsilon+i\infty} 
dp \exp(pt) g_m(p)~~~,
\label{doll} 
\end{equation}
\end{subequations} 
with $\epsilon>0$ an infinitesimal positive constant.  

Taking the Laplace transform of Eqs.~(\ref{thirteenmulti}a,b), and using 
the initial conditions
$E[C_{s_a}(0)]=C_{s_a}(0)=\delta_{aA}$ and 
$E[C_m(0)]=C_m(0)=0, m\not\in \{s_a\}$, we get
\begin{subequations}
\label{fifteenmulti}
\begin{eqnarray}
pg_{s_a}(p)-\delta_{aA}&=&-{1\over 8} \sigma^2  \sum_b (V^2)_{{s_a}{s_b}} 
g_{s_b}(p) \nonumber \\
&+&\sum_{n \not\in \{s_a\} }
 (-i) V_{s_an}f_n g_n(p+iE_n-iE_s)~~~,
\label{anise}  
\end{eqnarray}
and for $m\not \in \{s_a\}$,   
\begin{eqnarray}
pg_m(p)&=& -{1\over 8}\sigma^2(E_m-E_s)^2) g_m(p)  \nonumber \\ 
&+& \sum_b(-i)V_{ms_b}f_m g_{s_b}(p+iE_s-iE_m)~~~.
\label{bay} 
\end{eqnarray}
\end{subequations}

Solving Eq.~(\ref{bay}) for $g_m(p)$,  $m\not \in \{s_a\}$, and shifting 
$p \to p+iE_m$ in the solution, we get  
\begin{subequations}
\label{sixteenmulti}
\begin{equation}
g_m(p+iE_m)=[p+iE_m+{1\over 8}\sigma^2(E_m-E_s)^2]^{-1}
\sum_b(-i)V_{ms_b}f_m g_{s_b}(p+iE_s)~~~.
\label{aorta}   
\end{equation}
Shifting $p \to p+iE_s$ in Eq.~(\ref{anise}), and then substituting Eq.~(\ref{aorta}), 
we get an algebraic equation for the set of quantities $g_{s_b}(p+iE_s)$, 
\begin{eqnarray}
\sum_b K_{ab} g_{s_b}(p+iE_s)&=&\delta_{aA}~~~, \nonumber \\
K_{ab}= (p+iE_s)\delta_{ab}&+&{1\over 8}\sigma^2 (V^2)_{{s_a}{s_b}} 
+\sum_{m\not \in \{s_a\}} { f_m^2 V_{s_am}V_{ms_b}  
\over p+iE_s+i(E_m-E_s)f_m}~~~.
\label{beat}  
\end{eqnarray}
\end{subequations}

In physically interpreting these equations, we must remember that the 
Laplace transform variable $p$ is related to the usual energy variable 
$E$ by $p=-iE$.  Making this substitution in Eqs.~(\ref{sixteenmulti}a,b) we have 
respectively
\begin{subequations}
\label{seventeenmulti}
\begin{equation}
g_m(-iE+iE_m)=[-iE+iE_m+{1\over 8}\sigma^2(E_m-E_s)^2]^{-1}
\sum_b(-i)V_{ms_b}f_m g_{s_b}(-iE+iE_s)~~~,
\label{aussie} 
\end{equation}
and
\begin{eqnarray}
&&\sum_b K_{ab} g_{s_b}(-iE+iE_s)=\delta_{aA}~~~, \nonumber \\
&&K_{ab}= (-iE+iE_s)\delta_{ab} 
+{1\over 8}\sigma^2 (V^2)_{{s_a}{s_b}} 
+\sum_{m\not \in \{s_a\}} { f_m^2 V_{s_am}V_{ms_b}  
\over -iE+iE_s+i(E_m-E_s)f_m}~~~.
\label{boora}  
\end{eqnarray}
Corresponding to the changes of variable that have been made, the 
inversion formulas become 
\begin{eqnarray}
E[C_{s_a}(t)]&=&{1\over 2\pi} \exp(iE_st) 
\int_{i\epsilon-\infty}^{i\epsilon+\infty} dE
\exp(-iEt)g_{s_a}(-iE+iE_s) ~~~,\nonumber \\
E[C_{m\not\in \{s_a\}}(t)]&=&{1\over 2\pi} \exp(iE_mt) 
\int_{i\epsilon-\infty}^{i\epsilon+\infty} dE
\exp(-iEt)g_m(-iE+iE_m) ~~~.
\label{cheeky}  
\end{eqnarray}
\end{subequations}  

Inspecting the equation for the kernel $K_{ab}$, we see that apart from 
order $V^2$ terms it is a diagonal matrix $(-iE+iE_s) \delta_{ab}$.  Hence 
the solution $g_{s_b}(-iE+iE_s)$, on the inversion contour of integration,  
will be appreciable only in the vicinity 
of $E=i\epsilon + E_s$, that is, only near energy shell.  This motivates the 
Weisskopf-Wigner approximation of replacing $E$ in the denominator 
of the final  
term in $K_{ab}$ by $i\epsilon +E_s$, with the result 
that $K_{ab}$ then becomes a 
linear function of $E$.  Before making this approximation, the kernel 
$K_{ab}$ has a non-trivial dependence on the stochasticity 
parameter $\sigma$.  However, after making the Weisskopf-Wigner 
approximation, this $\sigma$ dependence completely cancels:
\begin{subequations}
\label{eighteenmulti}
\begin{eqnarray}
&&{1\over 8}\sigma^2 (V^2)_{{s_a}{s_b}} 
+\sum_{m\not \in \{s_a\}} { f_m^2 V_{s_am}V_{ms_b}  
\over -iE+iE_s+i(E_m-E_s)f_m}  \nonumber \\
\rightarrow&&{1\over 8}\sigma^2 (V^2)_{{s_a}{s_b}} 
+\sum_{m\not \in \{s_a\}} { f_m^2 V_{s_am}V_{ms_b}  
\over\epsilon+i (E_m-E_s)f_m}  \nonumber \\
=&&{1\over 8}\sigma^2 (V^2)_{{s_a}{s_b}} 
+\sum_{m\not \in \{s_a\}} { f_m V_{s_am}V_{ms_b}  
\over\epsilon+ i(E_m-E_s)} \nonumber \\
=&&{1\over 8}\sigma^2 (V^2)_{{s_a}{s_b}} 
+\sum_{m\not \in \{s_a\}} { [1-(i/8)\sigma^2(E_m-E_s)] V_{s_am}V_{ms_b}  
\over\epsilon+ i(E_m-E_s)} \nonumber \\
=&&\sum_{m\not \in \{s_a\}} {-i  V_{s_am}V_{ms_b}  
\over E_m-E_s-i\epsilon}~~~, 
\label{amish} 
\end{eqnarray}
where in the final step we have made use of the condition of Eq.~(\ref{acrid}).  
Thus in the Weisskopf-Wigner approximation, the kernel $K_{ab}$ appearing 
in Eq.~(\ref{boora}) simplifies to 
\begin{eqnarray}
K_{ab}&=& (-iE+iE_s)\delta_{ab}  
+\sum_{m\not \in \{s_a\}} {-i  V_{s_am}V_{ms_b}  
\over E_m-E_s-i\epsilon}\nonumber \\
&=& (-iE+iE_s)\delta_{ab} +iM_{ab}+{1\over 2} \Gamma_{ab}~~~,\nonumber \\
M_{ab}&=&\sum_{m\not \in \{s_a\}}P {  V_{s_am}V_{ms_b}  
\over E_s-E_m}~~~,\nonumber \\
\Gamma_{ab}&=&
2\pi\sum_{m\not \in \{s_a\}}  V_{s_am}V_{ms_b} \delta(E_m-E_s)~~~, 
\label{basic}  
\end{eqnarray}
\end{subequations}
with $P$ in the definition of the ``mass matrix'' $M_{ab}$ the principal 
value.  These are the same as the formulas 
for the kernel in the absence of the 
stochastic terms in the Schr\"odinger equation.  
Thus, in the Weisskopf-Wigner approximation, the 
solution for $E[C_{s_a}(t)]$ is unmodified by 
the stochastic effects, and hence the Lorentzian line profile and the 
decay rate of the state are unaffected by the $\sigma$ terms.  

The solution for $E[C_m(t)]$ with $m \not \in \{s_a\}$ does retain a  
dependence on the stochastic parameter.  To study this, let us specialize 
to the case $D=1$ of a non-degenerate initial state.   The expression 
in Eq.~(\ref{basic}) for the kernel now becomes the $1 \times 1$ matrix
\begin{subequations}
\label{nineteenmulti}
\begin{equation}
K(E)=-iE+iE_s +iM+{1\over 2} \Gamma~~~,
\label{arbor} 
\end{equation} 
with $M$ and $\Gamma$ real numbers given by 
\begin{eqnarray}
M&=& \sum_{m\neq s} P {  V_{sm}V_{ms}  
\over E_s-E_m}~~~,\nonumber \\
\Gamma&=& 2\pi\sum_{m\neq s}  V_{sm}V_{ms} \delta(E_m-E_s)~~~. 
\label{barn}  
\end{eqnarray}
Thus,  Eq.~(\ref{boora}) has the immediate solution
\begin{equation} 
g_s(-iE+iE_s)=K(E)^{-1}~~~,
\label{chill}  
\end{equation}
which when substituted into Eq.~(\ref{aussie}) yields   
\begin{equation}
g_m(-iE+iE_m)=[-iE+iE_m+{1\over 8}\sigma^2(E_m-E_s)^2]^{-1}
V_{ms}f_m (E-E_s-M+{i\over 2}\Gamma)^{-1}~~~.
\label{drink}  
\end{equation}
\end{subequations}

Substituting these equations into the inversion formulas of Eqs.~(\ref{cheeky}), 
and doing elementary contour integrations, we find 
\begin{subequations}
\label{twentymulti}
\begin{eqnarray}
\label{apex} 
E[C_s(t)]&=&\exp(-iMt-{1\over 2}\Gamma t), \\
E[C_{m\neq s}(t)]&=&{ V_{ms} \over E_s-E_m+M-{i\over 2}\Gamma}
\left(\exp[i(E_m-E_s-M)t-{1\over 2}\Gamma t]
-\exp[-{1\over 8}\sigma^2(E_m-E_s)^2t]\right)~.\nonumber
\end{eqnarray}
>From Eq.~(\ref{brazil}) we thus get
\begin{equation} 
E[|C_s(t)|^2]=\exp(-\Gamma t)~~~,
\label{bottom}   
\end{equation}
which identifies $\Gamma$ as the transition rate per unit time out of 
the initial state.  
Finally, substituting Eq.~(\ref{apex}) into Eq.~(\ref{brooch}), simplifying 
to leading order 
in $V$, and integrating with respect 
to $t$, we get 
\begin{eqnarray}
E[|C_{m\neq s}(t)|^2]&=&
{|V_{ms}|^2 \over (E_s-E_m+M)^2 +{1\over 4}\Gamma^2}
\biggl( \exp(-\Gamma t) +1   \nonumber \\
&  & \mbox{}-2 \exp[-{1\over 8}\sigma^2(E_m-E_s)^2 t-{1\over 2} \Gamma t] 
\cos[(E_s-E_m+M)t] \biggr)~~~.
\label{center}  
\end{eqnarray}
This completes our solution for the expectations of the coefficients, and  
their squared magnitudes, 
in the case of a non-degenerate initial state. We see that after a time 
$t$ large compared with the lifetime $\Gamma^{-1}$, we obtain 
\begin{equation}
E[|C_{m\neq s}(\infty)|^2]=
{|V_{ms}|^2 \over (E_s-E_m+M)^2 +{1\over 4}\Gamma^2}~~~,
\label{depth}   
\end{equation}
\end{subequations}
exhibiting the standard Lorentzian profile with no dependence on the 
stochasticity parameter $\sigma$.

\section{\label{gold}Small Time and Golden Rule Approximations}

Let us now study the behavior of Eq.~(\ref{center}) for small and large values of 
the time $t$.  Since within our approximations we have 
$\sigma^2 (E_m-E_s)^2 \simeq \sigma^2 [(E_s-E_m+M)^2 +{1\over 4} \Gamma^2]$, 
we can rewrite Eq.~(\ref{center}) as
\begin{subequations}
\label{twentyonemulti}
\begin{eqnarray}
\label{aplomb} 
E[|C_{m\neq s}(t)|^2]&=&
{|V_{ms}|^2 \over (E_s-E_m+M)^2 +{1\over 4}\Gamma^2}
\biggl( \exp(-\Gamma t) +1   \\
& & \mbox{} -2 \exp[-{1\over 8}\sigma^2 \big((E_s-E_m+M)^2 
+{1\over 4} \Gamma^2\big) t
-{1\over 2} \Gamma t ] 
\cos[(E_s-E_m+M)t] \biggr)~~~.\nonumber
\end{eqnarray}
In the limit as $t \to 0$, we can develop the exponential 
and cosine functions in power series expansions, giving
\begin{equation}
E[|C_{m\neq s}(t)|^2]\simeq |V_{ms}|^2 \left( {1\over 4} \sigma^2 t + t^2 
+ O(t^2\sigma^4(E_s-E_m)^2) +  O(t^3) \right)~~~.
\label{bedlam}  
\end{equation}
Thus the leading small time behavior of the summed 
expected probability in the 
decay channels is
\begin{equation} 
\sum_{m \neq s} |V_{ms}|^2  {1\over 4} \sigma^2 t 
=(V^2)_{ss} {1\over 4} \sigma^2 t~~~,
\label{carefree}  
\end{equation}
\end{subequations} 
where in evaluating the sum we have employed the condition of Eq.~(\ref{acrid}).  
We shall verify this result by another method in Sec. \ref{qze}, where we discuss 
its implications for the quantum Zeno effect, and in Sec. \ref{khakki} shall apply it   
to estimating bounds on $\sigma$.  

Let us next consider the large time behavior implied by Eq.~(\ref{aplomb}).  Once 
$t$ is large enough so that $|(E_m-E_s)t|$ is large for all energies 
$E_m$ not infinitesimally close to $E_s$, we can evaluate the summed expected 
probability in the decay channels by making the ``golden rule'' 
approximation \cite{multi}.  This approximation treats the factors 
multiplying $|V_{ms}|^2$ in Eq.~(\ref{aplomb}), which are sharply peaked around 
$E_m=E_s$, as if they were equal to a Dirac delta function of strength 
given by the integral of these factors over energy.  
We then have
\begin{subequations}
\label{twotwomulti}
\begin{eqnarray}
\sum_{m\neq s} E[|C_{m\neq s}(t)|^2]   
&=&\sum_{m\neq s}
{|V_{ms}|^2 \over (E_s-E_m+M)^2 +{1\over 4}\Gamma^2}
\biggl( \exp(-\Gamma t) +1    \nonumber \\
&-&2 \exp[-{1\over 8}\sigma^2 \big((E_s-E_m+M)^2 +{1\over 4} \Gamma^2\big) t
-{1\over 2} \Gamma t] 
\cos[(E_s-E_m+M)t] \biggr) \nonumber \\
&\simeq &\sum_{m\neq s}  |V_{ms}|^2 \delta(E_m-E_s)\nonumber \\ 
&\times&\int_{-\infty}^{\infty} 
d(\Delta E)  { \exp(-\Gamma t) +1
-2 \exp[-{1\over 8}\sigma^2 \big((\Delta E )^2 +{1\over 4} \Gamma^2\big) t
-{1\over 2} \Gamma t] 
\cos[(\Delta E) t]  \over   (\Delta E)^2 +{1\over 4}\Gamma^2} \nonumber \\
&=& \sum_{m\neq s}  |V_{ms}|^2 \delta(E_m-E_s) t F[\sigma^2/(8t),t]~~~\nonumber \\
&=&{\Gamma t \over 2 \pi}  F[\sigma^2/(8t),t]~~~,
\label{armor}  
\end{eqnarray}
with $\Gamma$ as given in Eq.~(\ref{barn}) and with 
the function $F[A,t]$ defined by 
\begin{equation}
F[A,t]= \int_{-\infty}^{\infty} du 
 { \exp(-\Gamma t) +1
-2 \exp[-A (u^2 +{1\over 4} \Gamma^2 t^2) 
-{1\over 2} \Gamma t] 
\cos u \over u^2 +{1\over 4}\Gamma^2t^2}~~~.
\label{armored} 
\end{equation}
\end{subequations}
To evaluate $F[A,t]$ we note that \cite{gradsh}
\begin{subequations}
\label{twentythreemulti}
\begin{equation}
F[0,t]= \int_{-\infty}^{\infty} du 
 { \exp(-\Gamma t) +1
-2 \exp(-{1\over 2} \Gamma t) 
\cos u \over   u^2 +{1\over 4}\Gamma^2t^2}
={2 \pi \over \Gamma t} [1-\exp(-\Gamma t)] ~~~,
\label{autumn} 
\end{equation}
and \cite{gradsh} 
\begin{eqnarray}
{\partial F[A,t] \over \partial A}&=&\int_{-\infty}^{\infty} du
2 \exp[-A (u^2 +{1\over 4} \Gamma^2 t^2) 
-{1\over 2} \Gamma t] \cos u  \nonumber \\
&=&2 \pi^{1/2}\exp(-{1\over 2}\Gamma t-{1\over 4} A \Gamma^2 t^2)  
A^{-1/2} \exp[-1/(4A)] ~~~.
\label{beauty}    
\end{eqnarray}
Thus, integrating Eq.~(\ref{beauty}) with respect to $A$ we get
\begin{equation}  
F[A,t]= {2 \pi \over \Gamma t} [1-\exp(-\Gamma t)] + C[A,t]~~~,
\label{collect} 
\end{equation}
\end{subequations}   
with the correction term $C[A,t]$ given by
\begin{subequations}
\label{twentyfourmulti}
\begin{equation}
C[A,t]=4 \pi^{1/2} \exp(-{1\over 2}\Gamma t)  
\int_0^{A^{1/2} } dv \exp[-{1\over 4} (v^2 \Gamma^2 t^2+ 1/v^2) ]~~~.
\label{aphid}  
\end{equation}
Since the exponentials of negative arguments in Eq.~(\ref{aphid}) are bounded 
by their maxima over the range of integration, we have 
\begin{equation}
|C[A,t]| < 4 \pi^{1/2} A^{1/2}\exp[-1/(4A)] 
= 2 \sigma ({ \pi \over 2t})^{1/2}  \exp(-2t/\sigma^2) ~~~.
\label{beetle}
\end{equation}
So when $\Gamma t$ is of order unity, the contribution 
of the correction term $C[A,t]$ is of order $c\sigma \Gamma^{1/2}  
\exp[-2/(\sigma^2 \Gamma)]$ $\sim c^{\prime} \sigma V \exp[-c^{\prime\prime}
/(\sigma V)^2]$, with $c,c^{\prime},c^{\prime\prime}$ constants, 
which is exponentially small and can 
be neglected in our approximation scheme.  
Thus we are justified in approximating 
\begin{equation}
F[A,t] \simeq F[0,t] = {2 \pi \over \Gamma t} [1-\exp(-\Gamma t)] ~~~,
\label{cocoon}   
\end{equation}   
which when substituted back into Eq.~(\ref{armor}) gives
\begin{equation} 
\sum_{m\neq s} E[|C_{m\neq s}(t)|^2] =  1-\exp(-\Gamma t) =1-|C_s(t)|^2~~~,
\label{dragonfly}   
\end{equation}
\end{subequations}
verifying that the approximations used in our calculation are 
consistent with maintenance of the unitarity sum rule (the unit state   
vector normalization condition).  

\section{\label{juxta}Solution to the Stochastic Equation for $C_{m\neq s}$}

Since we see from Eqs.~(\ref{apex}) and (\ref{center}) that $E[|C_{m\neq s}(t)|^2$ 
differs from $|E[C_{m\neq s}(t)]|^2$, the stochastic fluctuations in 
$C_{m\neq s}(t)$ are evidently playing a role.  Let us now demonstrate this 
directly by solving the stochastic differential equation for 
$C_{m\neq s}(t)$.  Specializing to the case of a non-degenerate initial 
state, approximating $C_s(t) \simeq E[C_s(t)]$, and using 
Eq.~(\ref{apex}) for  $E[C_s(t)]$, Eq.~(\ref{devotee}) becomes 
\begin{subequations}
\label{twentyfivemulti}
\begin{eqnarray}
dC_m(t)&=&(\alpha_m^{(1)} dW_t + \alpha_m^{(2)} dt) C_m(t)  \nonumber \\ 
&+&\exp[i(E_m-E_s-M)t-{1\over 2} \Gamma t] 
(\gamma^{(1)}_{ms} dW_t+\gamma^{(2)}_{ms}dt)~~~,\nonumber \\
\alpha_m^{(1)}&=&{1\over 2} \sigma (E_m-E_s)~,~~
\alpha_m ^{(2)}=-{1\over 8}\sigma^2(E_m-E_s)^2=-{1\over 2} (\alpha_m^{(1)})^2~~~,\nonumber \\
\gamma_{ms}^{(1)}&=&{1\over 2}\sigma V_{ms}~,~~
\gamma_{ms}^{(2)}\simeq-iV_{ms}f_m~~~.
\label{atrium}   
\end{eqnarray}
For general values of 
the coefficients $\alpha_m^{(1,2)}$ and $\gamma_{ms}^{(1,2)}$,   
Eq.~(\ref{atrium}) can be integrated by using Eqs.~(\ref{twomulti}a-c) to 
find a stochastic integrating factor for the $C_m$ 
terms (see the Appendix), with the 
result
\begin{eqnarray}
C_m(t)&=&\exp[\alpha_m^{(1)} W_t-(\alpha_m^{(1)})^2t]
\int_0^t \exp[i(E_m-E_s-M)u-{1\over 2} \Gamma u 
-\alpha_m^{(1)} W_u+(\alpha_m^{(1)})^2u]  \nonumber \\
& \times & [\gamma_{ms}^{(1)} dW_u +(\gamma_{ms}^{(2)}-\alpha_m^{(1)} 
\gamma_{ms}^{(1)})du]~~~.
\label{boxwood}  
\end{eqnarray}
Using Eqs.~(\ref{twomulti}a-c), it is easy to verify directly that Eq.~(\ref{boxwood}) solves 
Eq.~(\ref{atrium}).  If we now examine Eq.~(\ref{boxwood}) more closely, using the specific 
expressions for the coefficients $\alpha_m^{(1,2)}$ 
and $\gamma_{ms}^{(1,2)}$ given in Eq.~(\ref{atrium}), we find that within the  
approximation of neglecting terms of relative order $\sigma V^2$,  
the integrand   
in Eq.~(\ref{boxwood}) is an exact stochastic differential. Thus the integration 
can be carried out explicitly (see the Appendix), with the result
\begin{eqnarray}
\label{calla} 
C_{m \neq s}(t)&=&{V_{ms}\over E_s-E_m+M-{i\over 2}\Gamma}  \\
&\times& \left( \exp[i(E_m-E_s-M)t-{1\over 2} \Gamma t]  
-\exp[{1\over 2} \sigma(E_m-E_s)W_t-{1\over 4} \sigma^2(E_m-E_s)^2 t]\right).\nonumber
\end{eqnarray}
This expression can be easily verified, by use of Eqs.~(2a-c), to be the 
solution to Eq.~(\ref{atrium})   
(up to a residual error of relative order $\sigma V^2$).  
Using Eq.~(\ref{chi}) to take the expectation of 
Eq.~(\ref{calla}), we recover the result 
of Eq.~(\ref{apex}).  From Eq.~(\ref{calla}) we find an explicit formula for 
$|C_{m \neq s}(t)|^2$, 
\begin{eqnarray}
|C_{m \neq s}(t)|^2 &=&{|V_{ms}|^2 \over (E_s-E_m+M)^2+{1\over 4}\Gamma^2 }\nonumber \\
&\times& \left(\exp(-\Gamma t) 
 +\exp[ \sigma(E_m-E_s)W_t-{1\over 2} \sigma^2(E_m-E_s)^2 t] \right.\\
& -&\left. 2 \exp[{1\over 2} \sigma(E_m-E_s)W_t
-{1\over 4} \sigma^2(E_m-E_s)^2 t -{1\over 2} \Gamma t] 
\cos(E_m-E_s-M)t \right).\nonumber
\label{daisy}  
\end{eqnarray}
\end{subequations}
Again using Eq.~(\ref{chi}) to take the expectation of this formula, we recover  
the result of Eq.~(\ref{center}).

\section{\label{qze}Stochastic Suppression of the Quantum Zeno Effect}

In Eqs.~(\ref{twentyonemulti}b,c) we saw that the leading small time behavior of the 
summed expected probability in the 
decay channels is
\begin{equation} 
(V^2)_{ss} {1\over 4} \sigma^2 t~~~,
\label{wassail}   
\end{equation}
rather than the result $(V^2)_{ss}  t^2$ that would hold for  
vanishing $\sigma$.  As a result, $E[|C_s(t)|^2]-1$ vanishes linearly 
in $t$ for nonzero $\sigma$, rather than quadratically in $t$ as for 
the unmodified Schr\"odinger equation. Since the quadratic vanishing 
of $|C_s(t)|^2-1$ in standard quantum mechanics is the origin of the 
quantum Zeno effect \cite{misra}, we conclude that in the energy driven 
stochastic Schr\"odinger equation, the quantum Zeno effect is suppressed.  

Let us verify this directly from the stochastic differential equation of 
Eq.~(\ref{amber}), in analogy with the direct calculation \cite{anan} of $|C_s(t)|^2-1$ for 
small times for the ordinary Schr\"odinger equation. Applying the It\^o  
rule of Eq.~(\ref{almond}), we have 
\begin{subequations}
\label{twentysevenmulti}
\begin{equation}
d|\langle s(0)|s(t) \rangle |^2 |_{t=0}  
=\langle s(0)| d|s(t)\rangle |_{t=0} + \langle s(0)| d|s(t)\rangle^* |_{t=0}
+ \langle s(0)| d|s(t)\rangle |_{t=0} \langle s(0)| d|s(t)\rangle^* |_{t=0}~~~.  
\label{athlete} 
\end{equation}
{}From Eq.~(\ref{amber}) we have
\begin{equation} 
d|s(t)\rangle=-iH|s(t)\rangle dt - {1\over 8} \sigma^2
(H-\langle s(t)| H |s(t) \rangle)^2 |s(t)\rangle dt + {1\over 2} \sigma 
(H-\langle s(t)| H |s(t) \rangle) |s(t) \rangle  dW_t~~~,
\label{baseball} 
\end{equation}
and so setting $t=0$ and projecting on $\langle s(0)|$ gives
\begin{equation} 
\langle s(0)| d|s(t) \rangle|_{t=0} = 
-i\langle H \rangle_s dt - {1\over 8} \sigma^2 
\langle (H  -\langle H \rangle_s)^2 \rangle_sdt ~~~,
\label{cheer} 
\end{equation}
\end{subequations}
with $\langle H^n \rangle_s =\langle s(0) | H^n | s(0) \rangle$.  
Substituting Eq.~(\ref{cheer}) into Eq.~(\ref{athlete}), we thus get the first term in the 
small $t$ expansion of $ |\langle s(0)|s(t) \rangle |^2-1$, 
\begin{subequations}
\label{twoeightmulti}
\begin{equation}
|\langle s(0)|s(t) \rangle |^2=1-{1\over 4} \sigma^2 
\langle (H-\langle H \rangle_s)^2 \rangle_s t + O(t^2)~~~.
\label{attic} 
\end{equation}
This equation gives a general formula for the stochastic suppression of the 
quantum Zeno effect, independent of any assumptions about the potential. 
When the general form \cite{anan} of the order $t^2$ term coming from the  
standard Schr\"odinger evolution is included, Eq.~(\ref{attic}) becomes 
\begin{equation}
|\langle s(0)|s(t) \rangle |^2=1- 
\langle (H-\langle H \rangle_s)^2 \rangle_s  
\left( {1\over 4} \sigma^2 t + t^2 \right) 
+O(\sigma^4 t^2) + O(t^3)~~~;
\label{boxes}   
\end{equation}
in other words, the first two terms in the small $t$ expansion are governed 
to leading order in $\sigma$ by the initial state energy variance.  
When the potential is assumed to obey Eq.~(\ref{acrid}), we have 
\begin{eqnarray}
\langle H \rangle _s &=& E_s + V_{ss} = E_s~~~,\nonumber \\ 
\langle H^2 \rangle _s &=& E_s^2 + 2E_s V_{ss} + (V^2)_{ss}~~~,\nonumber \\
\langle (H-\langle H \rangle_s)^2 \rangle_s&=& 
\langle H^2 \rangle_s -\langle H \rangle_s^2 =(V^2)_{ss}~~~, 
\label{clutter}   
\end{eqnarray}
and so Eq.~(\ref{boxes}) becomes
\begin{equation} 
|\langle s(0)|s(t) \rangle |^2=1-(V^2)_{ss} 
\left( {1\over 4} \sigma^2 t + t^2 \right) +O(\sigma^4 t^2)  + O(t^3)~~~,
\label{discard}  
\end{equation}
\end{subequations}
in agreement with the result of Eqs.~(\ref{twentyonemulti}a,b) and the unitarity sum rule.

\section{\label{khakki}Discussion and Estimates of Bounds on $\sigma$}

We have seen that to leading order in the perturbing potential, the 
stochastic terms governed by $\sigma$ do not affect either the Lorentzian 
line profile or the transition rate per unit time as evaluated in the 
Weisskopf-Wigner approximation, but only produce a change 
in the short time transient behavior of the transition probabilities from 
the initial state.  This is a direct result of the fact that the 
energy-driven stochastic Schr\"odinger equation is energy conserving.  
On dimensional grounds, the transition rate per unit time 
$\Gamma$ could contain, 
in addition to the usual terms of the form $\delta(E_s-E_m)|V_{ms}|^2$, a  
term of the form $\sigma^2 (V^2)_{ss}$.  However, this additional term 
is not energy conserving, and as a result we have seen that its coefficient 
precisely cancels to zero in the Weisskopf-Wigner approximation.  

Because the transition rate per unit time and Lorentzian line shape are 
unaffected by $\sigma$, bounds 
on $\sigma$ from particle decays result only from experiments in which a 
metastable system is monitored as function of time from a known time 
(or vertex location) of formation.  According to Eqs.~(\ref{twoeightmulti}a-d), for small 
times the effective transition rate per unit time is 
\begin{subequations}
\label{twoninemulti}
\begin{equation}
\Gamma_R = {1\over 4} \sigma^2 (\Delta E)^2={1\over 4} \sigma^2 (V^2)_{ss}~~~,
\label{apostle}  
\end{equation}
with $ (\Delta E)^2= \langle (H-\langle H \rangle_s)^2 \rangle_s$ 
the initial state energy 
variance.  This can be interpreted as an early time decay 
rate coming from spontaneous reduction induced by the stochastic fluctuation 
terms, in agreement with the estimate $\Gamma_R \sim \sigma^2 (\Delta E)^2$
used in earlier discussions \cite{gisin,adler}.  In order for the rate of Eq.~(\ref{apostle}) to 
not lead to pronounced early time deviations from the 
observed decay rate $\Gamma$, we must have
\begin{equation} 
\Gamma_R < \Gamma~~~,
\label{believe}   
\end{equation}
\end{subequations} 
which writing $\sigma^2=M_{\sigma}^{-1}$ implies the bound 
\begin{equation}
M_{\sigma} > {(V^2)_{ss} \over 4 \Gamma}
={\sum_{m\not=s} |V_{sm}|^2 \over 
8\pi \sum_{m\not=s} |V_{sm}|^2 \delta(E_m-E_s)}
\equiv {E_D \over 8 \pi} ~~~,
\label{zeta} 
\end{equation}
with $E_D$ defining an energy characteristic of the decay process.  
In a particle 
physics context, a first guess would be to estimate $E_D$ as being      
of order the mass of the decaying particle.      
The most massive decays for which  $\Gamma$ has been measured 
by tracking a metastable system from the point of formation appear to be 
$\pi^0\to \gamma \gamma$ decay, with an initial mass order 140 MeV,  
and charmed meson decays, with an initial mass of around 2 GeV.     
Estimating $E_D$ in Eq.~(\ref{zeta}) as the decaying particle mass, 
these give respective bounds on $M_{\sigma}$ of  
order 6 MeV and 80 MeV, respectively.  If $M_{\sigma}$ were significantly larger 
than these bounds, one would have observed anomalous accumulations of decay 
events close to the production vertex, as a result of decays induced by 
spontaneous reduction.  For comparison, the observation of 
coherent superpositions of energy eigenstates in the neutrino, $K$-meson 
and $B$-meson systems gives bounds \cite{sladler}, respectively, of $M_{\sigma} >10^{-20}$GeV, 
$M_{\sigma}>2 \times 10^{-15}$GeV, and $M_{\sigma} >2 \times 10^{-13}$GeV.   

Thus the charmed meson decay bound on $M_{\sigma}$ 
represents a significant improvement over the coherent 
oscillation bounds.  However, it  
is still smaller than the Planck mass, which is very likely the expected 
value of $M_{\sigma}$, by a factor of $10^{20}$!  We conclude that the theory of  
decaying states in the energy-driven stochastic Schr\"odinger equation 
places only very weak empirical bounds on the magnitude of the stochasticity 
parameter $\sigma$.  

We leave for future study two issues that can be 
addressed within the general framework established here.  The first is 
an analysis of the 
nature of the transition between the short-time regime with decay rate 
$\Gamma_R$, and the exponential decay regime with decay rate $\Gamma$.  
This is governed by the solution of Eqs.~(\ref{seventeenmulti}a,b)  before making the 
Weisskopf-Wigner approximation of replacing $E$ in the denominator of the 
final term in $K_{ab}$ by $E_s$.  The second is an analysis of the magnitude of the 
energy $E_D$  
defined by Eq.~(\ref{zeta}), for various dynamical models of the decay process, as 
reflected in the energy spectrum of the unperturbed states $|m\rangle$ and 
in the magnitudes of the decay-inducing matrix elements $V_{sm}$.  

\section{\label{stem}Acknowledgments}

This work was supported in part by the Department of Energy under
Grant \#DE--FG02--90ER40542.   I wish to acknowledge the hospitality of 
the Aspen Center for Physics, where most of the calculations reported 
here were done.  I also 
wish to thank Angelo Bassi, Dorje Brody, Todd Brun, and 
especially Lane Hughston,  
for informative discussions 
about the properties of the It\^o calculus,  Todd Brun for a helpful 
discussion about extracting bounds from decay observations,  
Larry Horwitz for a 
discussion several years ago about virtues of the Laplace transform, 
and Edward    
Witten for raising the issue of 
empirical bounds on the stochasticity 
parameter.  


\section{\label{adn}Added Note}

Lajos Di\'osi \cite{ldiosi} has pointed out an elegant stochastic-theoretic 
technique that 
allows the main physical results of this paper to be derived in a few lines,  
starting from the standard quantum mechanical results that hold when 
the stochasticity parameter $\sigma$ is zero.  
Di\'osi makes three principal  
observations. The first is 
that the quantities of direct physical interest, as pointed out in Sec. \ref{juror}, 
are the expectations  $E[|C_m(t)|^2]$ of the squared magnitudes of the 
perturbation coefficients. Since according to Eq.~(\ref{betsy}) we have
\begin{subequations}
\label{threeonemulti}  
\begin{equation}
|C_m(t)|^2 = |\langle m|\psi(t)\rangle|^2  =\langle m|\psi(t)\rangle 
\langle \psi(t) |m \rangle   =\langle m|\rho(t) |m \rangle~~~,
\label{waif}  
\end{equation}
with $\rho(t)$ the density matrix
\begin{equation} 
\rho(t)=|\psi(t) \rangle \langle \psi(t)| ~~~,
\label{waltz}    
\end{equation}  
to calculate $E[|C_m(t)|^2$ it suffices to know $E[\rho(t)]$, in other words 
\begin{equation}
E[|C_m(t)|^2]= \langle m| E[\rho(t)]|m \rangle 
= {\rm Tr}  (|m \rangle \langle m|) E[\rho(t)]~~~.
\label{waive}   
\end{equation}
Since the dynamics of $E[\rho(t)]$  is governed by the Lindblad-type equation 
of Eq.~(\ref{deep}), to calculate the physically relevant expectations it thus 
suffices to solve the dynamical problem specified by Eq.~(\ref{deep}), supplemented  
by the initial condition
\begin{equation}
E[\rho(0)]= \rho(0)=|s_A\rangle \langle s_A|~~~.
\label{wary}   
\end{equation}
\end{subequations}

Di\'osi's second observation is that the dynamical problem specified by 
Eq.~(\ref{deep}), with the initial condition of Eq.~(\ref{wary}), can be compactly 
solved by a simple stochastic trick.  The trick uses  
the fact \cite{noise} that there is a second stochastic     
Schr\"odinger equation, simpler in structure than that of Eq.~(\ref{amber}), which 
also leads to Eq.~(\ref{deep}) as the equation for the evolution of the stochastic  
expectation of its density matrix.  To see this, consider the stochastic 
Schr\"odinger equation
\begin{subequations}
\label{threetwomulti}
\begin{equation}
d|\psi\rangle=-iH|\psi\rangle dt -{1\over 8} \sigma^2  H^2 |\psi\rangle dt
+{1\over 2} i \sigma H |\psi\rangle dW_t~~~.
\label{yoga}  
\end{equation}
This equation differs from that of Eq.~(\ref{amber}) in having an imaginary noise 
term, with operator coefficient $H$, instead of a real noise term with 
operator coefficient $H-\langle H \rangle$.  
A simple calculation, using the It\^o calculus rules of Eq.~(\ref{beige}), 
shows that Eq.~(\ref{yoga}) also leads to preservation of the norm of the state 
$|\psi\rangle$, and leads to the density matrix evolution equation
\begin{equation}
d\rho=i[\rho,H]dt-{1\over 8}\sigma^2[H,[H,\rho]]dt +{1\over 2} i \sigma  
[H,\rho] dW_t~~~,
\label{yogi}   
\end{equation}
which has the stochastic expectation
\begin{equation} 
dE[\rho]=i[E[\rho],H]dt -{1\over 8} \sigma^2 [H,[H,E[\rho]]]dt ~~~,
\label{yucca}  
\end{equation}
\end{subequations}
which is identical to Eq.~(\ref{deep}).  Hence the imaginary noise equation of 
Eq.~(\ref{yoga}) will lead to the same results for the physical quantities 
$E[|C_m(t)|^2]$ as the real noise equation of Eq.~(\ref{amber}), even though the 
stochastic details of the two processes differ!  

Di\'osi's third observation is the fact that  Eq.~(\ref{yoga}) can be immediately 
formally integrated to give
\begin{subequations}
\label{thirtythreemulti}
\begin{equation}
|\psi(t)\rangle = \exp[-iH(t - {1\over 2}\sigma W_t)] |\psi(0\rangle~~~,
\label{zebra}   
\end{equation} 
as can be readily ascertained by use of Eq.~(\ref{cocoa}) with the choice
\begin{equation} 
\alpha={1\over 2} i \sigma H~~~.
\label{zany} 
\end{equation}
\end{subequations}
Combining this observation with the first two,  
then leads to a very simple rule for calculating the   
stochastic modifications of decay processes governed by Eq.~(\ref{amber}).  
Let $E[|C_m^{\sigma}(t)|^2]$ be the quantities of physical interest, 
viewed as functions of $\sigma$ as well as of $t$, so that 
$E[|C_m^{0}(t)|^2]=|C_m^{0}(t)|^2 $  are their values as calculated 
from the standard 
Schr\"odinger evolution with no stochasticity.  Then Eqs.~(\ref{waif}) through 
(33b) imply the simple relation
\begin{subequations}
\label{thirtyfourmulti}
\begin{equation} 
E[|C_m^{\sigma}(t)|^2]  = E[|C_m^{0}(t-{1\over 2}\sigma W_t)|^2]~~~,
\label{rankle}    
\end{equation} 
between the probabilities calculated in the standard Schr\"odinger analysis, 
and the stochastic expectations of the probabilities as calculated from 
Eq.~(\ref{amber}).  The recipe is simply this:  take the known expressions for 
the probabilities calculated in  standard quantum mechanics, replace 
$t$ by $t-{1\over 2} \sigma W_t$, and take the stochastic expectation.  
The needed stochastic expectations of powers of $W_t$ can all be read off 
from the expansion of Eq.~(\ref{chi}) in powers of $\alpha$,
\begin{equation} 
E[W_t]=0~,~~E[W_t^2]=t~,E[W_t^3]=0~,~~E[W_t^4]=3t^2~,... ~~.
\label{ransack}   
\end{equation}
\end{subequations}

Let us now apply Di\'osi's observations to rederive the principal results 
found above for the stochastic analog of the Weisskopf-Wigner analysis.     
First, let us consider the short time behavior of the 
survival probability given in Eq.~(\ref{boxes}).  The standard answer when  
$\sigma=0$, which gives the quantum Zeno effect, is
\begin{subequations}
\label{threefivemulti}
\begin{equation}
|\langle s(0)|s(t) \rangle|^2= 1-\langle (H-\langle H \rangle_s)^2
\rangle_s t^2 + O(t^3)~~~.
\label{recant}     
\end{equation}
Following the recipe, we have
\begin{equation} 
E[(t-{1\over 2}\sigma W_t)^2]=E[t^2-t \sigma W_t 
+ {1\over 4} \sigma^2 W_t^2] = t^2 +{1\over 4} \sigma^2 t~~~.
\label{reflex}   
\end{equation}
On substitution into Eq.~(\ref{recant}) this gives for general $\sigma$ 
\begin{equation}
|\langle s(0)|s(t) \rangle|^2= 1-\langle (H-\langle H \rangle_s)^2
\rangle_s  (t^2+{1\over 4} \sigma^2 t)  +... , 
\label{reflux}    
\end{equation}
\end{subequations}
in agreement with the result for the stochastic modification of the 
quantum Zeno effect given in Eq.~(\ref{boxes}).  

Next let us apply the recipe to the formula for the initial state survival 
probability obtained using the Weisskopf-Wigner approximation, which is 
valid for times $t$ that are not too small (and also not too large).  
The standard analysis gives
\begin{subequations}
\label{threesixmulti}
\begin{equation}
 |C_s^{0}(t)|^2  = \exp(-\Gamma t)~~~,
\label{elba}    
\end{equation}  
with $\Gamma$ the Golden Rule decay rate of Eq.~(\ref{barn}).  Replacing 
$t$ by $t-{1\over 2} \sigma W_t$ and using Eq.~(\ref{chi}) to take the  
stochastic expectation, we get as the exact formula for the stochastic 
modification of the Weisskopf-Wigner approximation 
\begin{equation}
E[|C_s^{\sigma}(t)|^2]  = \exp [
-\Gamma (1-{1\over 8} \sigma^2 \Gamma ) t ]~~~,
\label{egis}   
\end{equation}  
which reduces, when the  correction term of relative order  
$\sigma^2 \Gamma$ is neglected, to 
the answer found in Eq.~(\ref{bottom}).  
Since $\sigma^2 \Gamma \sim O(\sigma^2 V^2)$, we see that 
the calculation of Secs. 4-6 above did not succeed in keeping 
all terms of order $\sigma^2 V^2$,  and in fact there is  
a small stochastic 
correction to the decay rate, with the corrected decay rate given by  
\begin{equation}
\Gamma^{\sigma}=\Gamma (1-{1\over 8} \sigma^2 \Gamma )~~~.
\label{eked}   
\end{equation}
\end{subequations}
However, writing $\sigma^2=M_{\sigma}^{-1}$ as in Sec. \ref{khakki}, as long as $M_{\sigma} >E_s$  
this correction is not significant within 
the Weisskopf-Wigner approximation scheme, which treats the line width   
$\Gamma$ as a small quantity relative to $E_s$.    
                
Finally, let us apply the recipe to the formula giving the probability for 
a transition to the state $|m\rangle$.  The standard 
Weisskopf-Wigner approximation result for this is given by Eq.~(\ref{center}) with 
$\sigma=0$,
\begin{subequations}
\label{threesevenmulti} 
\begin{eqnarray}
|C_{m\neq s}^{0}(t)|^2&=&
{|V_{ms}|^2 \over (E_s-E_m+M)^2 +{1\over 4}\Gamma^2}
\biggl( \exp(-\Gamma t) +1   \nonumber \\
&& \mbox{} -2 \exp[-{1\over 2} \Gamma t] 
\cos[(E_s-E_m+M)t] \biggr)~~~.
\label{frieze}   
\end{eqnarray}
Applying the recipe, and again using Eq.~(\ref{chi}) to evaluate the needed 
expectations, we get the exact stochastic extension of Eq.~(\ref{frieze}), 
\begin{eqnarray}
E[|C_{m\neq s}^{\sigma}(t)|^2]&=&
{|V_{ms}|^2 \over (E_s-E_m+M)^2 +{1\over 4}\Gamma^2}
\biggl( \exp[-\Gamma(1 -{1\over 8} \sigma^2 \Gamma) t ] +1   \nonumber \\
&& \mbox{} -2 \exp[-{1\over 2} \Gamma
( 1-{1\over 16}\sigma^2\Gamma)t-{1\over 8} \sigma^2 (E_s-E_m+M)^2 t ]  \nonumber \\
&\times&\cos[(E_s-E_m+M)(1-{1\over 8} \sigma^2\Gamma)t] \biggr)~~~.
\label{tactile}   
\end{eqnarray}
\end{subequations}
Again, when simplified to leading order in $V$, this gives the result of  
Eq.~(\ref{center}) above.  However, even before dropping nonleading 
terms in $V$, we see that Eq.~(\ref{tactile}) implies the Lorentzian formula 
of Eq.~(\ref{depth}) in the large time limit.  

{}From the above exposition, we see that Di\'osi's observations not only 
greatly simplify the calculation of the physically relevant quantities, but 
also give results that are completely independent of the assumption of 
Eq.~(\ref{acrid}) that was used to linearize the stochastic equation.  (This is 
something that one might have already suspected from the fact 
that Eq.~(\ref{boxes}) is more general 
than Eq.~(\ref{discard}).)  Thus, the only approximations that are needed to get 
stochastic results are those that are used 
in the standard, non-stochastic quantum mechanical analysis.  
Moreover, the ``miraculous'' 
cancellation of the $\sigma^2$ terms in the Weisskopf-Wigner approximation 
to the mass and decay matrices, 
exhibited above in Eq.~(\ref{amish}), is given a deeper explanation.    
There is an extensive literature \cite{facchi} 
discussing the decay problem without making the Weisskopf-Wigner 
approximation (i.e., without replacing $E$ by $E_s$ in the order $V^2$ terms   
of the Laplace transform kernel $K_{ab}$), and these discussions 
can all be converted to results for $E[|C_m^{\sigma}(t)|^2]$ 
in the stochastic case, by using the recipe of replacing $t$ 
by $t-{1\over 2} \sigma W_t$ in the corresponding formula for $|C_m^0(t)|^2$ 
and taking a stochastic average over $W_t$.    


The relation of Eq.~(\ref{rankle}) between stochastic and standard quantum 
mechanical probabilities can be applied to other problems as well.  
For example, the density matrix of a two-level system can be 
represented in the form
\begin{subequations}
\label{threeeightmulti}
\begin{equation} 
\rho ={1\over 2} (1-\vec R \cdot \vec \tau)~,
\label{intat} 
\end{equation}
with $\vec \tau=(\tau_1,\tau_2,\tau_3)$ the standard Pauli matrices, and  
with $\vec R=(R_1,R_2,R_3)$ a vector summarizing the structure of the 
traceless part of the density matrix.  
The standard, $\sigma=0$ Schr\"odinger equation describing Rabi oscillations 
of the two-level system under the influence of an applied field oscillating 
at the frequency of the level separation (in co-rotating coordinates, 
neglecting the counter-rotating field component) 
gives for the equation 
of motion \cite{iirabi} of the vector $\vec{R}|_{\sigma =0} \equiv \vec{R}^0$, 
\begin{equation}
{d \vec R^0 \over dt}= \vec \omega \times \vec R^0~,
\label{intale}    
\end{equation}
with $|\vec \omega|=\Omega$ the angular frequency of precession of 
$\vec R^0$. The probabilities for finding the system in the 
upper and lower 
levels are  given, as a function of time, by
\begin{equation} 
P_{\pm}^0(t)={1\over 2} [1\pm R_3^0(t)]~.
\label{intart}   
\end{equation}
\end{subequations}

Since the general solution of Eq.~(\ref{intale}) has the form 
\begin{subequations}
\label{threeninemulti}
\begin{equation}
\vec R^0(t)= \vec V_1 \cos \Omega t + 
\vec V_2 \sin \Omega t~,   
\label{thtea}  
\end{equation}
with $\vec{V}_{1,2}$ fixed vectors that depend on the initial state and the 
structure of the Hamiltonian, and since 
\begin{eqnarray}
E[\cos \Omega (t-{1\over 2}\sigma W_t) ]&=&
 \exp(-{1\over 8}\Omega^2 \sigma^2 t)  \cos \Omega t   ~,\nonumber \\
E[\sin \Omega (t-{1\over 2}\sigma W_t) ]&=&
 \exp(-{1\over 8}\Omega^2 \sigma^2 t) \sin \Omega t  ~,
\label{thtap}   
\end{eqnarray}
we have under the stochastic evolution of Eq.~(\ref{amber})
\begin{equation}
E[\vec R^{\sigma}(t)]= \exp(-{1\over 8}\Omega^2 \sigma^2 t)\vec R^0(t)   ~.
\label{thtax}    
\end{equation}
By Eq.~(\ref{intart}), this gives for the expected probabilities when the system 
evolves under the stochastic Schr\"odinger equation, 
\begin{equation}
{1\over 2} - E[P_{\pm}^{\sigma}(t)]
=  \exp(-{1\over 8}\Omega^2 \sigma^2 t)  [{1\over 2} - P_{\pm}^0(t)]~.
\label{thex}   
\end{equation}
\end{subequations}
This can be applied, for example, to the quantum Zeno effect experiment 
of Itano et. al. \cite{itano}, who carry out a proposal of Cook \cite{cook} to make 
repeated measurements of a two-level system while the vector $\vec R$ is 
precessing for a time interval $t=\pi/\Omega$, for which the exponential   
damping factor in Eq.~(\ref{thtax}) becomes $\exp(-{1\over 8} \pi \Omega \sigma^2)$.
Corresponding to the experimental value  $\Omega=320.7$ MHz and the 
fact that probabilities were observed to an accuracy of about $.02$ in 
this experiment, and were found to agree with the standard Schr\"odinger 
theory, we get a bound on 
$M_{\sigma}=1/\sigma^2$ of $M_{\sigma} > 2 \times 10^{-15}$ 
GeV, comparable to that obtained from oscillations in the $K$-meson 
system.  


\appendix*
\section{}

We give here the details of the integration of the stochastic differential 
equation that appears in Sec. \ref{juxta}.  Consider the linear stochastic differential 
equation 
\begin{equation}
dC_t=(A_t dW_t + B_t dt) C_t +P_t dW_t + Q_t dt~~~,
\label{apply}   
\end{equation} 
which is to be solved for the unknown stochastic 
function $C_t$ given the known functions 
$A_t$, $B_t$, $P_t$, and $Q_t$.  Although we shall proceed as if these   
known functions were deterministic, in fact all our manipulations and 
the final solution are unchanged \cite{hughston} if the input functions are themselves  
stochastic.  To solve Eq.~(\ref{apply}), we transpose the $C_t$ term on the 
right to the left and multiply 
by a factor $F_t$, which is to be determined, giving  
\begin{equation}
F_t[dC_t-(A_t dW_t + B_t dt) C_t]  =F_t[P_t dW_t + Q_t dt ]~~~.
\label{apduo}   
\end{equation} 
We now look for an $F_t$ which makes the left hand side of Eq.~(\ref{apduo}) a 
total differential, up to terms independent of $C_t$ that are 
of the same form as the 
terms on the right hand side.  Making the Ansatz
\begin{equation} 
F_t = \exp\big[\int_0^t (\alpha_u dW_u + \beta_u du)\big]~~~,
\label{aptrio}  
\end{equation}
we find by use of Eqs.~(\ref{almond}) and (\ref{cocoa}) of the text that
\begin{equation} 
d(F_t C_t)=F_t[dC_t+\alpha_t dW_t C_t
+(\beta_t +{1\over 2} \alpha_t^2) dt C_t 
 + \alpha_t dW_t dC_t]~~~,
\label{apquad}  
\end{equation} 
which on substituting Eq.~(\ref{apply}) for the final $dC_t$ on the right, and 
using Eq.~(\ref{beige}) of the text, gives
\begin{subequations}
\label{cincomulti}
\begin{equation} 
d(F_t C_t)=F_t[dC_t +\alpha_t dW_t C_t  
+(\beta_t +{1\over 2} \alpha_t^2+\alpha_t A_t) dt C_t 
+  \alpha_t P_t dt]~~~.
\label{apquint}   
\end{equation} 
Hence if we choose
\begin{equation} 
\alpha_t = -A_t~,~~~\beta_t=-B_t  + {1\over 2} A_t^2~~~,
\label{apquintb}    
\end{equation}
\end{subequations}
then Eq.~(\ref{apquint}) takes the form
\begin{equation} 
d(F_t C_t)=F_t[dC_t -(A_t dW_t +B_t dt) C_t  
-A_t P_t dt]~~~,
\label{apsextet}  
\end{equation} 
which by use of Eq.~(\ref{apduo}) becomes
\begin{equation} 
d(F_t C_t) = F_t[P_t dW_t + (Q_t -A_t P_t) dt]~~~.
\label{apseptet}  
\end{equation}   
The dependence on the unknown function $C_t$ is now entirely in the form 
of an exact differential, and so Eq.~(\ref{apseptet}) can be immediately integrated 
to give 
\begin{eqnarray}
\label{apoct} 
C_t&=&F_t^{-1}\big[C_0+\int_0^t du F_u \big(P_u dW_u +(Q_u-A_u P_u)du 
\big) \big] \nonumber \\
&=&\exp\big[\int_0^t \big( A_u dW_u + (B_u-{1\over 2} A_u^2) du  
\big)\big] \\
&\times&\left( C_0 + \int_0^t du 
\exp\big[-\int_0^u \big( A_v dW_v + (B_v-{1\over 2} A_v^2) dv  \big)\big] 
[P_u dW_u +(Q_u-A_u P_u) du] \right) ,\nonumber
\end{eqnarray}
which is the general solution of Eq.~(\ref{apply}).

In Sec. \ref{juxta}, we need only the case of Eq.~(\ref{apply}) in which $A_t=A$, 
$B_t =B$, $P_t =P f_t$, and $Q_t=Q f_t$, with $A,B,P,Q$ constants and 
with $f_t$ of the form $f_t=\exp(Kt)$, and so the solution of Eq.~(\ref{apoct}) then   
becomes 
\begin{eqnarray}
C_t&=&\exp \big( A W_t + (B-{1\over 2} A^2) t  \big) \nonumber \\
&\times&\left( C_0 + \int_0^t du 
\exp \big( -A W_u + (K-B+{1\over 2} A^2) u \big) 
[P dW_u +(Q-A P) du] \right) .
\label{apan}   
\end{eqnarray}
Using the identity (proved by the same methods used to find the integrating 
factor $F_t$), 
\begin{eqnarray}
&~&\exp\big(\alpha W_u+(\beta +K)u\big) [PdW_u+(Q+\alpha P) du] \\
&=&{P\over \alpha} d\exp\big(\alpha W_u+(\beta +K)u\big)
-{P\over \alpha} (\beta +K -{Q\over P} \alpha -{1\over 2} \alpha^2) 
\exp\big(\alpha W_u+(\beta +K)u\big)du , \nonumber
\label{apdec}  
\end{eqnarray}
and taking $\alpha=-A$ and $\beta=-B+A^2/2$, the $P dW_u$ term in Eq.~(\ref{apan}) 
can be eliminated.  This gives an alternate form for the solution $C_t$, 
\begin{eqnarray}
C_t&=&\exp \big( A W_t + (B-{1\over 2} A^2) t  \big) \nonumber \\
&\times&\left( C_0  
 -{P\over A}  \big[  \exp \big( -A W_t +(K-B+{1\over 2} A^2) t \big)
 -1 \big]   \right. \nonumber \\ 
&+&\left. \left[ {P\over A} (K-B)+Q \right]  
\int_0^t du \exp \big( -A W_u + (K-B+{1\over 2} A^2) u \big)  
 \right) ~~~.
\label{apele}   
\end{eqnarray}
Taking 
\begin{eqnarray}
A&=&\alpha_m^{(1)}={1\over 2} \sigma (E_m-E_s)~~~,\nonumber \\
B&=&-{1\over 2} (\alpha_m^{(1)})^2=-{1\over 8} \sigma^2 (E_m-E_s)^2~~~,\nonumber \\
P&=&{1\over 2} \sigma V_{ms}~~~,\nonumber \\
Q&=&-i V_{ms}f_m~,~~f_m=1-{i\over 8} \sigma^2(E_m-E_s)~~~, \nonumber \\
K&=&i(E_m-E_s-M) - {1\over 2} \Gamma~,
\label{apdoz}   
\end{eqnarray}     
in Eqs.~(\ref{apan}) and (\ref{apele}) gives the results quoted respectively in Eqs.~(\ref{boxwood}) 
and (\ref{calla}) of the text.  



\end{document}